# Wavelength and Polarization Multiplexed Nonlocal Metasurface for Quantitative Phase Microscopy

Haiwei Wang[1*], Shikun Ma[2,3], Shaban B. Sulejman[1], Niken Priscilla[1], Wendy S.L. Lee[1], Peter Francis Matthew Elango[4,5], Lukas Wesemann[1], Elizabeth Hinde[2] and Ann Roberts[1]

[1] ARC Centre of Excellence for Transformative Meta-Optical Systems, School of Physics, The University of Melbourne, Victoria 3010, Australia

[2] School of Physics, The University of Melbourne, Victoria 3010, Australia

[3] Department of Biochemistry and Pharmacology, The University of Melbourne, Victoria 3010, Australia

[4] Functional Materials and Microsystems Research Group and the Mico Nano Research Facility, RMIT University, Melbourne, 3001, Australia

[5] ARC Centre of Excellence for Transformative Meta-Optical Systems, RMIT University, Melbourne, 3001, Australia

*Corresponding author: haiweiw@student.unimelb.edu.au

## Abstract

Imaging transparent samples remains an ongoing challenge in the study of unstained biological cells and material samples. Widely used methods trade off system complexity, cost and bulk, computational efficiency and information content. Here we demonstrate the use of a nonlocal metasurface located in the object plane for obtaining single-shot, low-noise differential phase contrast images visualising phase gradients along orthogonal directions in a sample obtained at wavelengths of 613 nm and 656 nm. Furthermore, we show that these images are sufficient to calculate the quantitative phase introduced into the transmitted optical field by the sample. We find that the recovered phase of an optical field generated by a spatial light modulator is in good agreement with specified values. We also present information-rich differential phase contrast images of unstained

HeLa cells with the recovered phase excursion values consistent with the literature. Our results demonstrate the potential for metasurfaces as a platform for extracting information from an optical field for use in next-generation compact imaging systems with applications in medical diagnostics, biotechnology, and materials science.

# 1 Introduction

Phase microscopy is a powerful technique where variations in the phase introduced into an optical field by a sample can be observed[1]. Due to the sensitivity of the optical phase to minute variations in the sample's refractive index and/or thickness, phase microscopy permits the observation of otherwise low-contrast samples without staining or the introduction of fluorescent labels. As such, phase contrast microscopies have become indispensable tools in biology and materials science, with the two most popular techniques being Zernike phase contrast[2], and differential interference contrast microscopy (DIC)[3]. However, both methods have limitations. DIC visualizes phase gradients along only a single direction and Zernike phase contrast requires careful alignment and can produce artefacts. Furthermore, these techniques generate only qualitative phase images where the phase is not directly measured. Quantitative phase microscopy (QPM), on the other hand, is an imaging modality that provides a numerically evaluated phase map of the sample. Not only does QPM confer the key advantages of phase contrast microscopies, the quantitative phase data obtained can also be used in additional analyses. For example, the additional information can be used to obtain the dry mass of cells and can be used to study the growth of cells over time[4]. Existing interferometric[5–10] and non-interferometric[11–16] methods of QPM generally utilize complex and bulky systems and/or require capturing multiple images.

In the effort to miniaturize QPM systems, there is ongoing interest in the application of metasurfaces to phase imaging[17]. Recently, QPM has been demonstrated predominantly using wavefront-shaping local metasurfaces to obtain information for input into various phase recovery algorithms[18–21]. On the other hand, nonlocal metasurfaces that perform direct spatial frequency filtering[17,22–26] without the requirement for a 4*f* optical system or similar, have been demonstrated primarily for qualitative phase imaging, with only a few examples that demonstrated phase quantification experimentally[26,27]. These metasurfaces generate wavevector-dependent transmission (or reflection) with dispersive nonlocal resonances such as guided mode resonances[26,28,29], resonances in multilayer film stacks[30–32], Mie resonances[33,34], or quasi-bound-states-in-the-continuum[35,36]. Unlike wavefront-shaping metasurfaces, metasurface spatial frequency filters do not have inherent imaging capabilities and can be integrated into existing imaging systems by placing them after the object or in the image plane. Most nonlocal metasurfaces that have been experimentally realized exhibit a symmetric response about normal incidence, making them suitable only for the extraction of edge features[34,35,37,38], or for generating contrast similar to Zernike phase contrast[26], unless used with off-normal illumination[24]. The use of an asymmetric angular transmission response about normal incidence, on the other hand, permits both the magnitude and sign of the phase gradient to be determined, providing an avenue to QPM by obtaining the gradients of the phase and subsequent processing[19,27,39]. To obtain a robust quantitative phase image using this approach, however, gradient contrast along different directions within the sample is required, often leading to increased system size or complexity[14,19,27,40].

In this work, we propose and experimentally demonstrate the use of a single-layer nonlocal metasurface, multiplexing phase information in two circular polarizations and at two different wavelengths to perform QPM as illustrated in Figure 1. As part of this process, two single-shot differential phase contrast (DPC) images visualizing phase gradients along orthogonal directions are obtained, which serve as inputs to the phase retrieval algorithm. At each of the two operating wavelengths, the sign of the gradient contrast is reversed in the left circularly (LCP) and right circularly polarized (RCP) images obtained with our metasurface, which are obtained simultaneously using a polarization-sensitive camera. These two images combine to produce a DPC image with relatively low noise, and with background contrast and other artefacts removed. By changing the wavelength, the direction of the gradient contrast is switched between the horizontal and vertical directions, permitting the acquisition of DPC along two axes for robust quantitative phase recovery. The nonlocal metasurface used here is based on the concept of optical routing into resonant waveguide gratings[25,41,42]. Using our platform, we demonstrate QPM of phase targets generated using a spatial light modulator (SLM) as well as of unstained HeLa cells. This work opens opportunities for future developments in metasurface-based, ultracompact QPM systems that can be used in point-of-care and remote settings.

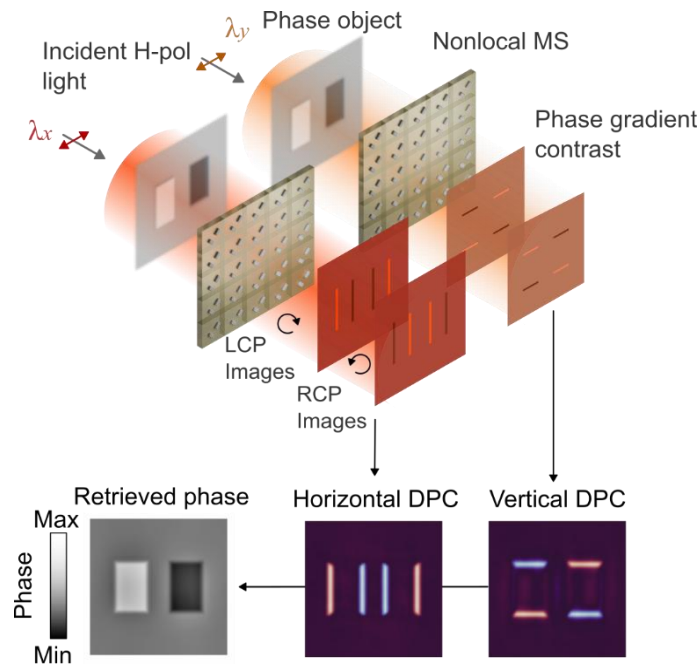

**Figure 1: Concept of spectral and polarization multiplexing-enabled quantitative phase imaging.** A phase object modelled as a thin, pure phase transmission mask is illuminated with horizontally polarized light at two different wavelengths. After the phase-modulated field is transmitted through the metasurface, phase gradient contrast images with opposing contrast directions appear in the LCP and RCP channels, and the phase gradient contrast along orthogonal directions at the two wavelengths channels. The images are used to construct horizontal and vertical gradient DPC images, which are then used to retrieve the phase.

## 2 Results

### 2.1 Theoretical framework

The objective of QPM is to determine the spatially varying phase, $\phi(\boldsymbol{\rho})$, a function of the transverse position $\boldsymbol{\rho} = (x, y)$, imprinted onto a transmitted field by a transparent sample. The sample is modelled as a thin, pure phase object with a transmission function

$$t(\boldsymbol{\rho}) = \exp[i\phi(\boldsymbol{\rho})]. \tag{1}$$

We assume that $\phi(\boldsymbol{\rho})$ is polarization-independent, and is given by

$$\phi(\boldsymbol{\rho}) = \frac{2\pi}{\lambda} \text{OPL}(\boldsymbol{\rho}), \tag{2}$$

where $\lambda$ is the free-space wavelength and $\text{OPL}(\boldsymbol{\rho})$ is the optical path length at a particular point on the sample. We assume that the $\text{OPL}(\boldsymbol{\rho})$ does not depend on wavelength such that $\phi(\boldsymbol{\rho})$ is inversely proportional to wavelength, permitting the incorporation of DPC images obtained at different wavelengths into the phase retrieval algorithm[43].

The sample is illuminated by a monochromatic, normally incident plane wave propagating along the $z$-axis with wavelength $\lambda$, amplitude $E_0$, and spatially uniform polarization given by the vector $\hat{\boldsymbol{e}}_{\text{in}}$. To describe the effect of a nonlocal metasurface, the spatial dependence of the field is decomposed into a sum of spatial frequencies specified by coordinates $\boldsymbol{u} = (u_x, u_y)$ with amplitudes calculated using the Fourier transform as defined in Supplementary Information 1. In free space, each value of $\boldsymbol{u}$ corresponds to a plane wave propagating at a distinct angle to the $z$-axis and incident on the metasurface. Assuming the incident light is scattered by only small angles by the object, the polarization upon scattering remains unchanged, and the Fourier spectrum of the field incident on the metasurface can be written as $\boldsymbol{E}_{\text{in}}(\boldsymbol{u}) = E_0 \tilde{t}(\boldsymbol{u}) \hat{\boldsymbol{e}}_{\text{in}}$, where $\tilde{t}(\boldsymbol{u}) = \mathcal{F}\{t(\boldsymbol{\rho})\}(\boldsymbol{u})$, and $\mathcal{F}$ is the Fourier transform operator as defined in Supplementary Information 1. A nonlocal metasurface with an angle dependent transmission modifies the spatial frequency spectrum of an incident field passing through it. This is characterized by a spatial frequency and wavelength dependent transmission tensor $\boldsymbol{T}_\lambda(\boldsymbol{u})$. After transmission through the metasurface and prior to detection, the field component with polarization given by $\hat{\boldsymbol{e}}_{\text{out}}$ is selected by means of a polarization analyzer. Finally, the measured intensity is proportional to the absolute value squared of this field:

$$I(\boldsymbol{\rho}; \lambda, \hat{\boldsymbol{e}}_{\text{in}}, \hat{\boldsymbol{e}}_{\text{out}}) = |E_0|^2 |\mathcal{F}^{-1}\{H(\boldsymbol{u}; \lambda, \hat{\boldsymbol{e}}_{\text{in}}, \hat{\boldsymbol{e}}_{\text{out}}) \tilde{t}(\boldsymbol{u})\}|^2, \tag{3}$$

where $\mathcal{F}^{-1}$ is the inverse Fourier transform, and $H(\boldsymbol{u}; \lambda, \hat{\boldsymbol{e}}_{\text{in}}, \hat{\boldsymbol{e}}_{\text{out}}) = \hat{\boldsymbol{e}}_{\text{out}} \cdot \boldsymbol{T}_\lambda(\boldsymbol{u}) \cdot \hat{\boldsymbol{e}}_{\text{in}}$ is defined as the coherent transfer function (CTF). We highlight here that different CTFs can be obtained for a metasurface with a fixed $\boldsymbol{T}_\lambda(\boldsymbol{u})$ by switching $\lambda$ and the input and output polarizations, enabling the transfer function to be selected externally without changing the metasurface or adding or removing components.

The DPC images are calculated from two intensity images with opposite gradient contrast $I_+(\boldsymbol{\rho})$, and $I_-(\boldsymbol{\rho})$, using:

$$I_{\text{DPC}}(\boldsymbol{\rho}) = \frac{I_+(\boldsymbol{\rho}) - I_-(\boldsymbol{\rho})}{I_+(\boldsymbol{\rho}) + I_-(\boldsymbol{\rho})}. \tag{4}$$

Here, these two images are obtained using a quarter waveplate in conjunction with a polarization-sensitive camera. In this case, the advantage of using this DPC signal in phase retrieval as opposed to raw images is that it is relatively low noise and removes background information. The former is due to subtraction of the two gradient-reversed images, where any variations common to both images (such as speckle) cancel, while the phase signal is reinforced. The latter benefit arises from normalizing by the sum of the two images, which removes the dependence on incident intensity, introduced, for example, by variations in absorption in the sample under investigation, permitting phase quantification without requiring a separate background (object-free) measurement.

Following the derivation presented in Supplementary Information 2, we show that the DPC signal depends only on the phase of the object through a convolution

$$\tilde{I}_{\text{DPC}}(\boldsymbol{u}) = H_{\text{DPC}}(\boldsymbol{u})\tilde{\phi}(\boldsymbol{u}), \tag{5}$$

where $\tilde{\phi}(\boldsymbol{u})$ and $\tilde{I}_{\text{DPC}}(\boldsymbol{u})$ are, respectively, the Fourier transforms of the phase and the DPC image, while $H_{\text{DPC}}(\boldsymbol{u})$ is the DPC transfer function, which can be calculated from the CTFs as detailed in Supplementary Information 2. Given a set of DPC signals as shown in Figure 1, the phase is then retrieved using noniterative deconvolution with Tikhonov regularization[44]:

$$\phi(\boldsymbol{\rho}) = \mathcal{F}^{-1}\left\{\frac{\sum_n H^*_{\text{DPC},n}(\boldsymbol{u})\tilde{I}_{\text{DPC},n}(\boldsymbol{u})}{\sum_n |H_{\text{DPC},n}(\boldsymbol{u})|^2 + \alpha}\right\}, \tag{6}$$

where $\tilde{I}_{DPC,n}(\boldsymbol{u})$, are the Fourier transforms of DPC signals each related to the phase via the DPC transfer function $H_{DPC,n}(\boldsymbol{u})$, and $\alpha$ is a regularization parameter selected to balance the requirements of minimizing noise while optimizing spatial resolution. Here, two DPC images $I_{\text{DPC},x}(\boldsymbol{\rho})$ and $I_{\text{DPC},y}(\boldsymbol{\rho})$ are captured, generated by respective DPC transfer functions $H_{DPC,x}(\boldsymbol{u})$ and $H_{DPC,y}(\boldsymbol{u})$, that provide both horizontal ($x$), and vertical ($y$) gradient contrast which ensures that a phase object is equally well-resolved along all directions in the image.

## 2.2 Metasurface design and characterization

Figure 2a shows a schematic of the unit cell of the metasurface, which consists of a periodic array of plasmonic nanorods on top of a silicon nitride waveguiding layer of thickness 150 nm. The metasurface is sandwiched between a 500 nm thick layer of polymethyl methacrylate (PMMA) as the superstrate, and a borosilicate glass substrate. The metallic nanorods, assumed rectangular, have lengths $l = 100$ nm, widths $w = 64$ nm and heights of $h = 33$ nm. To promote adhesion, a 3 nm layer of chromium was deposited prior to 30 nm of silver. The unit cell consists of two identical nanorods oriented perpendicularly to each other and positioned diagonally across the unit cell. The unit cell is rectangular with a periodicity of $P_x = 400$ nm along the $x$-axis and $P_y = 370$ nm along the $y$-axis. The center-to-center separation of the nanorods is $P_x/4$ along the $x$-direction and $P_y/4$ along the $y$-direction. The design was fabricated using a standard lithographic procedure and the details are outlined in the Methods section. The fabricated metasurfaces were of overall size 300 μm x 300 μm. Figure 2b shows scanning electron micrographs of a portion of the fabricated metasurface, showing a slight nonuniformity in the dimensions of the nanorods of different unit cells. This can cause random scattering of light which contributes to image noise.

The starting point for our design is the chiral guided-mode optical routing scheme[25,41,42] where the arrangement of two orthogonally oriented rectangular scatterers, as seen in Figure 2a, excites guided modes propagating along a single direction determined by the handedness of the incident illumination, resulting in approximately linear spatial dispersion about normal incidence. This contrasts with symmetric resonant waveguide gratings, which couple incident light into standing waves, resulting in parabolic spatial dispersion near normal incidence. As a consequence, the angle-dependent transmission at wavelengths near resonance exhibits a chirality-dependent asymmetry about normal incidence. This structure has been demonstrated previously to generate phase gradient contrast limited to a single axis since only modes propagating along one direction were excited[25]. The key modification in our design is to make the length of the unit cell in the $y$ direction comparable to that in the $x$ direction. As a result, guided modes propagating in the $y$ direction can be excited at a slightly shorter wavelength than those propagating in the $x$ direction. For the guided mode resonance at the shorter wavelength, we observe an asymmetric dispersion along $y$ instead of $x$, which permits changing the contrast direction by changing the incident wavelength.

Figure 2c-f shows a comparison of simulated and experimentally obtained angle-dependent transmission spectra with incident circularly polarized light and with no analyzer prior to detection. Numerical results were obtained with full-wave simulations using the finite element method implemented in COMSOL Multiphysics 6.3 with the Wave Optics module, while the experimental results were obtained using a custom benchtop spectroscopy setup. Further details of the simulation and spectroscopy setup are given in the Methods section. Three resonance features labelled 1-3 can be seen in both simulated and experimental results in Figure 2c-f. Note that the simulation results are plotted over a different wavelength range to the experimental results for a clearer comparison, since we found that the resonances observed in the experiments were blue-shifted compared to the simulations. With a 40 nm offset as shown, the experimental and simulated transmission spectra show good qualitative agreement. This discrepancy is likely due to fabrication intolerances and the choice of optical constants for the materials assumed in the simulations.

Figure 2c and e show the variation of the spectra for rotations corresponding to varying $u_y$ values, with $u_x$ held fixed at zero, where we see that feature 1 changes in response, while features 2 and 3 were largely unaffected. Figure 2d and f show the variation of the spectra for rotations corresponding to varying $u_x$ values, with $u_y$ held fixed at zero, where now we see that features 2 and 3 resonances now change in response to different incident angles. This result shows that feature 1 corresponds to the excitation of a guided mode that propagates along $y$, while features 2 and 3 correspond to the excitation of guided modes that propagate along $x$. This agrees with our expectation that the guided mode coupled along $y$ should occur at a shorter wavelength than the guided mode of the same order coupled along $x$. Furthermore, by reversing the handedness of the incident circularly polarized light, the spectra at positive angles mapped to those at negative angles, confirming the polarization routing behavior. This result is shown in Supplementary Figure 1. In the spectra, feature 2 can be attributed to a transverse-magnetic guided mode, while the other two features are associated with transverse-electric guided modes, deduced by applying the weakly perturbed slab waveguide model from Ref.[45].

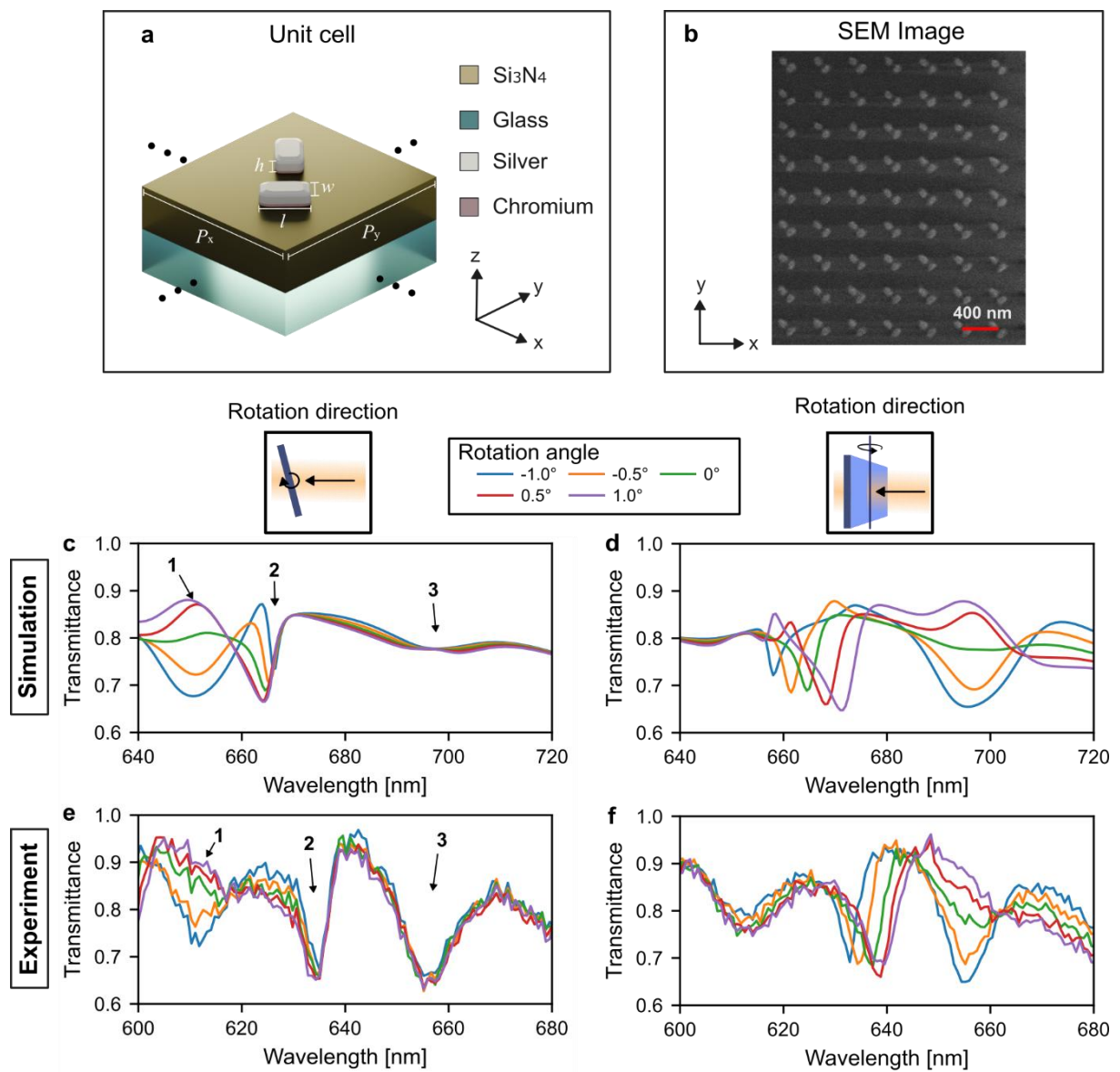

**Figure 2: Design and simulation of the metasurface.** (a) Schematic of a unit cell of the metasurface. (b) Scanning electron microscope (SEM) images of a portion of the device. (c, d) Simulated, and (e, f) experimentally measured transmission spectra at varying incident angles. (c, e)

Page | 9

tilting of the metasurface such that $u_y$ is varied while $u_x$ is held constant. (d, f) tilting of the metasurface such that $u_x$ is varied while $u_y$ is held constant.

Figure 3 shows the simulated and experimental magnitudes of the CTFs with incident linear polarization along $x$ and LCP and RCP output analyzers. As above, the simulations were performed using COMSOL Multiphysics 6.3 and the experimental results were obtained using a custom benchtop back focal plane imaging setup, both described in the Methods section. In contrast to the results obtained in Figure 2, where the metasurface was illuminated with circularly polarized light, we switched to the polarization multiplexing scheme using incident linearly polarized light and filtered output LCP and RCP light obtained using a quarter waveplate and a polarization sensitive camera for simultaneous acquisition of RCP and LCP images as shown in Figure 1. While the operating principle of the polarization-routing metasurface assumed incident circular polarization, we can use the fact that linearly polarized light consists of an even superposition of LCP and RCP light to obtain gradient contrast reversed images simultaneously. For the simulations, we observed vertical ($y$) contrast at $\lambda = 650$ nm, and horizontal ($x$) contrast at $\lambda = 695$ nm. For the experiments, we observed vertical ($y$) contrast at $\lambda = 613$ nm, and horizontal ($x$) contrast at $\lambda = 656$ nm. While the simulation results were obtained at a single wavelength, the experimental results used a quasi-monochromatic source with bandwidth of 6 nm full width at half-maximum.

Figure 3a and 3c show the simulated and experimentally obtained CTFs over a numerical aperture (NA) of 0.4, where similar features can be observed in both the simulation and experiment. Note here that the value of $|H|$ can exceed 1 due to polarization conversion. Guided mode resonances can be attributed to the sharp features observed in the plots due to their strong sensitivity to the angle of incidence. At the vertical DPC wavelength, two approximately horizontal curved features meet at the center, while at the horizontal DPC wavelength, two vertical curved features meet at the center. We see that one branch has lower transmission than the other, and this is switched upon changing the handedness of the circular polarization analyzer. Auxiliary features perpendicular to the main features are observed at oblique angles due to guided mode resonances coupled in the orthogonal direction.

Figure 3b and d show magnified plots of the same CTFs near normal incidence, showing the desired responses necessary for two-axis DPC. Due to the angular selectivity of guided mode resonances, this optical response is restricted to a relatively limited NA. The circled regions in Figure 3d indicate the effective numerical aperture of 0.05. Using this NA, the spatial resolution of our system was estimated to be 6.6 μm at 656 nm using Abbe's definition of resolution: $\lambda/(2NA)$. The maximum contrast of the transmittance values between positive and negative angles was measured to be 0.25 at 613 nm, and 0.4 at 656 nm. The simulated phase of the CTFs is shown in Supplementary Figure 2, where they are seen to be only slowly varying over the range of interest. While this was not confirmed experimentally, we can infer that the fabricated device also satisfies this condition from the correspondence between the other simulated and experimental results.

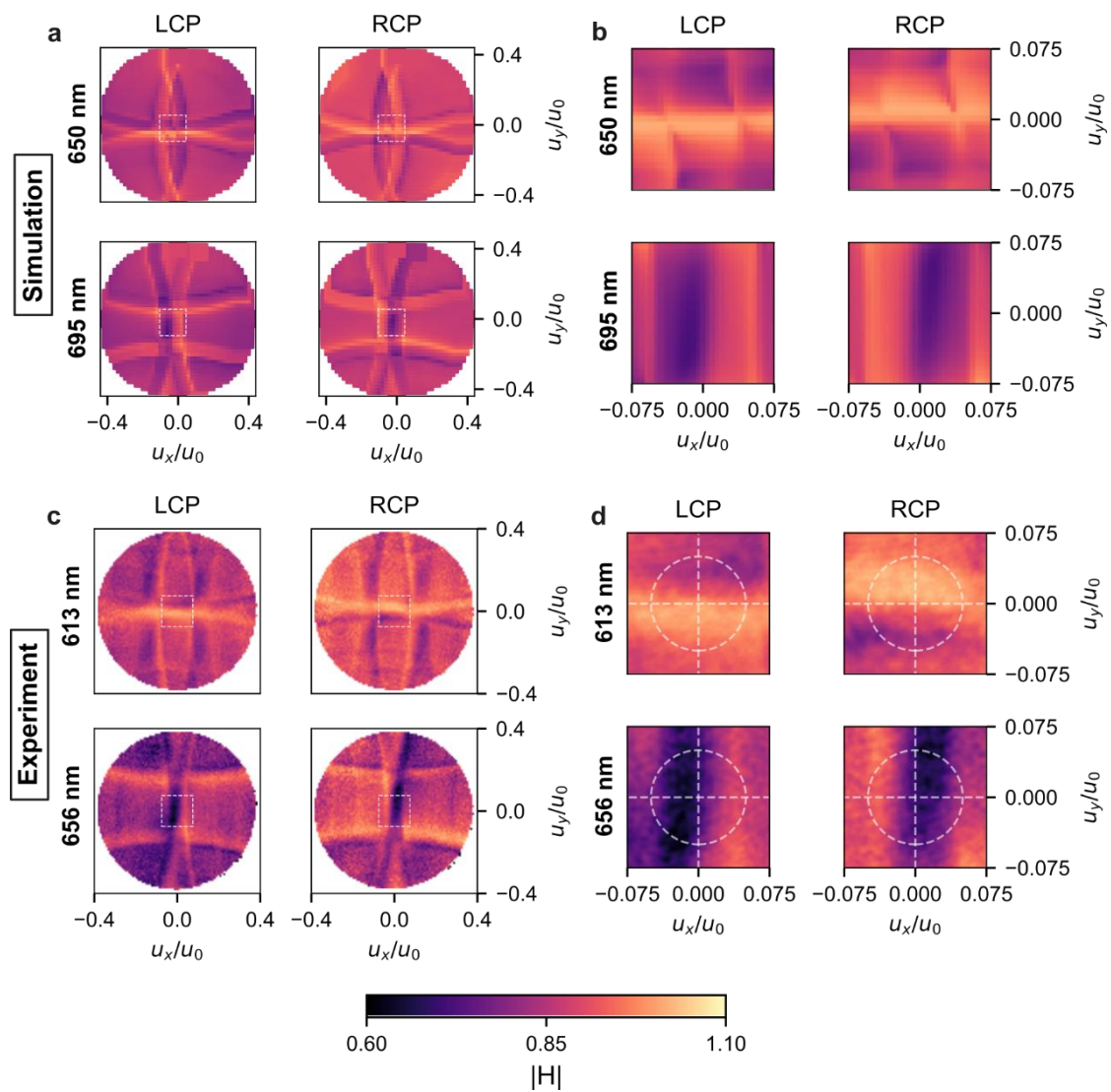

**Figure 3: Polarization and wavelength multiplexed coherent transfer functions.** (a, b) Simulated, and (c, d) experimentally measured absolute values of the CTFs of the metasurface for two operating wavelengths that give vertical (top row), and horizontal (bottom row) gradient contrast. (a, c) are plots over a large NA, while (b, d) are plots over a small NA, as indicated by the dashed squares in a and c. Here, the horizontal and vertical axes are normalized spatial frequencies, $u_x/u_0$ and $u_y/u_0$.

2.3 Demonstration of QPM

We first tested the quantitative phase imaging performance of our metasurface-based system using a reflective liquid crystal spatial light modulator (SLM) (Holoeye Pluto-VIS-001) to generate known phase distributions. The setup is described in the Methods section. To calibrate the phase values displayed by the SLM, we used the polarization-dependent properties of the SLM, detailed in Supplementary Information 4. The quasi-monochromatic light source and polarization setup was the same as that used for measuring the angle-dependence. We found that the OPL provided by the SLM is effectively independent of wavelength in the wavelength range considered, as evidenced by the result in Supplementary Figure 3, which shows that the SLM phase at 656 nm can be determined using the phase at 613 nm by applying Eq. 2. In the following, we report the phase at $\lambda = 656$ nm.

Figure 4 shows the results of quantitative phase imaging of sinusoidal star targets with varying phase excursions. The square regions shown correspond to a width of 233 μm at the plane of the metasurface. Figure 4a shows plots of the polarization images obtained from the LCP and RCP channels of the field obtained simultaneously using a polarization-sensitive camera, shown at the two operating wavelengths for a star target with a 1.5 rad phase excursion. Due to the background intensity noise, the modulations produced by the metasurface are difficult to perceive visually. Using the sum of the LCP and RCP channels, we calculate the background intensity $I_{background}(\boldsymbol{\rho}) = I_+(\boldsymbol{\rho}) + I_-(\boldsymbol{\rho})$, shown in Figure 4b. In the background images, the weak gradient features observed in Figure 4a are no longer prominent in Figure 4b, and any contrast of the object not due to the effects of the metasurface are isolated. Meanwhile, the DPC images shown in Figure 4c isolate the effects of

Page | 12

the metasurface, showing clear directional phase gradient contrast, without the common intensity variation between the two channels.

Figure 4d shows comparison of retrieved phase with the target phase projected from the SLM to validate our method. To determine the phase, the value of the regularization parameter was held constant at 0.005 and the experimentally measured CTFs were used to calculate the DPC transfer functions. The retrieved phase shows that each spoke in the image is approximately identical, revealing that our two-axis phase retrieval is isotropic. The lineplots show that the retrieved phase is consistent with the target phase for a range of spatial frequencies. For increasing phase excursions, the signal-to-noise ratio improves due to greater gradient contrast over the spatial noise.

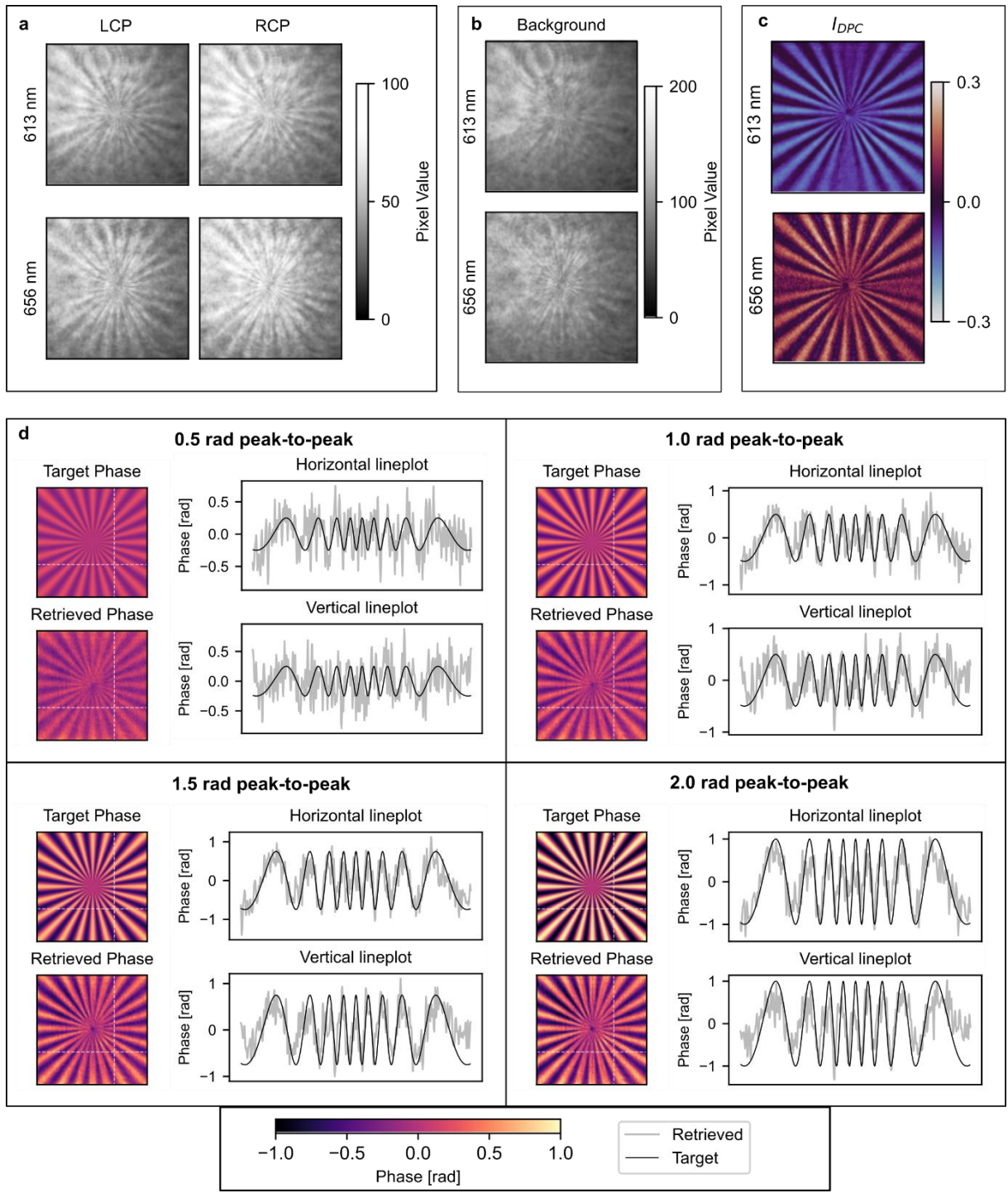

**Figure 4: Quantitative phase imaging of star targets.** (a) Filtered intensity measurements obtained with a quarter waveplate and a polarization-sensitive camera. (b) Background images obtained by summing images at the two polarizations. (c) DPC images obtained by subtracting the two polarizations and dividing by the background. (d) Comparison of the retrieved phase with the target phase projected on the SLM for the same target with varying phase excursions, as labelled in each cell. The square regions correspond to a physical size of $233 \times 233 \ \mu m^2$.

Finally, we applied our metasurface to imaging unstained biological cells, demonstrating its applicability to practical microscopy. We imaged HeLa (cervical cancer) cells grown on a coverslip, shown in Figure 5, with the preparation procedure outlined in the Methods section. A schematic showing the arrangement of components for imaging the cell samples is shown in Figure 5a. The sample was placed in the same setup used to image the SLM phase targets, with the SLM disabled such that it functioned as a flat mirror. The sample was positioned in the front focal plane of a microscope objective and the metasurface was placed immediately after the sample. Shifting the metasurface away from the object plane does not affect the spatial frequency response of the device. Figure 5b shows the retrieved phase, calculated with a regularization parameter of 0.005. The range of phase values is consistent with those given in the literature, which also use visible light with similar wavelengths as here[14,46]. Figure 5c shows the background and DPC images acquired in the process. Again, we see that while the background intensity contains coherent diffraction artifacts, the DPC images are mostly free of these artifacts. Figure 5d shows the comparison of a DIC image obtained with a commercial DIC microscope (Olympus BX60) in transmission mode with a 20x 0.4NA objective with a synthesized DIC image calculated using the retrieved phase. The synthesized DIC image was calculated by applying the transfer function of a DIC system with a diagonal shear distance of 5 μm to the field with a uniform intensity and phase given by the retrieved phase. We observe that while the DIC images maintain significant gradient contrast even for very small features such as the cell organelles, the DPC and phase images obtained with the metasurface have a lower spatial resolution due to the relatively low effective metasurface NA of 0.05.

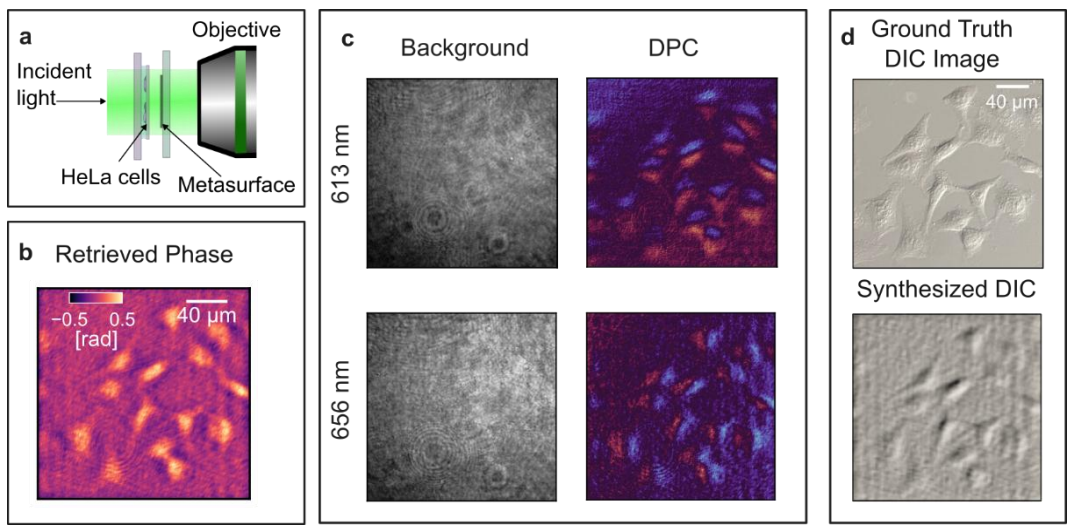

**Figure 5: HeLa cell imaging.** (a) Configuration for HeLa cell imaging. (b) Retrieved phase image. (c) Background and DPC images. (d) Comparison of an image acquired using a commercial DIC microscope and a synthesized DIC image using the retrieved phase in (a).

Discussion

We have proposed and experimentally demonstrated quantitative phase imaging using a nonlocal metasurface with wavelength- and polarization-multiplexed spatial filtering functions. Despite the spatially coherent illumination, the resulting DPC and retrieved phase images do not exhibit noise associated with speckle, as the image subtraction process eliminates the background. While polarization multiplexing enabled single-shot capture of two gradient-reversed images with a polarization-sensitive camera, wavelength switching was required to capture DPC along orthogonal axes in the image, which was speed restricted due to the software constraints of the tunable filter used with the supercontinuum laser source. Barring this limitation, the next bottleneck for QPM framerate is the speed of the phase retrieval algorithm, which requires only 3 discrete Fourier transform operations, and can be accelerated using faster processors or parallel computing techniques.

Future work will be aimed at obtaining measurements a single wavelength or integrating filters to permit single-shot wavelength multiplexed imaging as well as extending the metasurface NA. We do anticipate, however, that new approaches to engineering the metasurface transfer function will be required.

Conclusion

In conclusion, we have demonstrated a novel nonlocal metasurface engineered with four distinct spatial frequency responses accessed by tuning the wavelength and polarization for efficient quantitative phase contrast microscopy. This work opens the path to exploring additional functionalities that can be multiplexed onto a single layer metasurface. We envisage that such metasurfaces can be integrated with other meta-optical components and advanced light sources to produce imaging devices with unprecedented performance and compact form factors.

## 4 Methods

### 4.1 Simulation

Finite element method frequency domain simulations of the metasurface unit cell were conducted using COMSOL Multiphysics 6.3. A single rectangular unit cell was modelled, with Floquet periodic boundary conditions to emulate an infinite lattice and to model plane waves at off-normal incidences. Input and output port boundaries were placed sufficiently far from the metasurface to accommodate the near fields. The substrate and superstrate were modelled with fixed refractive indexes of 1.5, the refractive index of silicon nitride was modelled using a Sellmeier dispersion model with constants from Ref.[48], and silver and chromium were modelled with data for bulk materials given in Ref.[49]. At the output port, the complex amplitudes of transmitted LCP and RCP plane waves were computed. To calculate the spectra in Figure 2, the LCP and RCP transmittances were summed, while to calculate the CTFs in Figure 3, we calculated the S-parameters for LCP and RCP transmitted waves.

### 4.2 Fabrication

Silicon nitride was deposited onto borosilicate glass wafer pieces using plasma-enhanced chemical vapor deposition (Oxford Instruments Plasmalab 100 PECVD). A 200 nm thick layer of polymethyl methacrylate resist was spin-coated onto the substrate and baked for 3 minutes at 180°C followed by spin-coating a water-soluble conducting polymer (Dischem Discharge H2Ox2). The patterns for the metasurface were defined using electron beam lithography. The resist was developed using a 3:1 solution of IPA:MIBK. 3 nm thick chromium followed by deposition of a 30 nm thick silver was deposited using electron beam evaporation (Intlvac Nanochrome II), and the sample is placed in an acetone bath overnight for lift-off. A 500 nm thick PMMA cover layer was then spin-coated and baked to form a protective superstrate.

### 4.3 Optical characterization

The custom setups used for characterizing the angle-dependent and spectral transmission response (Figure 3) will be described briefly here. Diagrams of the systems and detailed descriptions of each component can be found in Supplementary Information 5.

To measure the transmission spectra at varying incident angles (Figure 3b and 3c), the sample was illuminated with collimated light from a multimode fiber coupled halogen lamp (OceanOptics HL-2000-HP). Before illuminating the sample, a linear polarizer (Thorlabs LPVIS050-MP2) and a quarter waveplate (Thorlabs AHWP10M-580) were used to circularly polarize the light. The sample was mounted on a motorized rotation stage (Thorlabs PRM1Z8) to control the tilt of the sample. The transmitted field of view was restricted by using an aperture, and the light was coupled into a visible light spectrometer (OceanOptics HR2000+ES) via a multimode optical fiber cable. A bright reference was taken using an unpatterned region of the sample.

To measure the absolute values of the CTFs (Figure 3d), we built a back focal plane imaging setup. The metasurface was illuminated with a focal spot generated by a high numerical aperture objective with its back aperture filled by a collimated, single-mode beam from a filtered supercontinuum source (NKT SuperK Compact with a SuperK Select acousto-optic filter). The beam was set to horizontal polarization using a linear polarizer (Thorlabs LPVIS050-MP2). An imaging objective collected the transmitted light, which then passed through a quarter waveplate (Thorlabs AHWP10M-580) with its fast axis at 45° to the metasurface axes, the back focal plane was relayed to a polarization-sensitive camera (Thorlabs Kiralux Polarization Camera CS505MUP1).

### 4.4 Imaging

The imaging setup used with the metasurface is described in detail and pictured in Supplementary Information 5. A brief description is given here. Collimated light from the filtered supercontinuum source given above was horizontally polarized and expanded using an objective and a lens to

Page | 18

illuminate the SLM (Holoeye Pluto-2). The beam, which may have phase imparted by the SLM was then demagnified and collimated using a lens and then an objective to illuminate the metasurface. During cell imaging, the SLM was disabled, and the cell sample was placed in the path of this narrow, collimated beam, which ensures sufficient light intensity across the region of interest. After the metasurface, the image of the SLM or sample was relayed to a polarization camera (given above) with an objective and a lens, passing through a quarter waveplate (Thorlabs AHWP10M-580) with its fast axis at 45° to the metasurface axes before the sensor.

4.5 Cell sample preparation

22 × 22 mm coverslips were placed in 35-mm imaging dish. Coverslips were coated with 0.1mg/mL Poly-D-Lysine (PDL) for 30 minutes at room temperature. HeLa cells were grown in DMEM (Lonza) supplemented with 10% bovine growth serum (Gibco), 1 × Pen-Strep (Lonza) at 37°C in 5% $CO_2$. Cells were fixed with 4% paraformaldehyde for 15 minutes at room temperature. After fixation, coverslips were transferred onto glass slides for imaging.

List of abbreviations

QPM – quantitative phase microscopy

DIC – differential interference contrast

DPC – differential phase contrast

LCP – left circularly polarized

RCP – right circularly polarized

OPLD – optical path length difference

CTF – coherent transfer function

PMMA – polymethyl methacrylate

NA – numerical aperture

SLM – spatial light modulator

Declarations

Availability of data and materials

The datasets used and/or analysed during the current study are available from the corresponding author on reasonable request.

Competing interests

A.R. serves as a member of the editorial board for the journal. The other authors declare that they have no competing interests.

Funding

This research was funded by the Australian Government through the Australian Research Council Centre of Excellence grant (CE200100010). E.H. acknowledges support from the Australian Research Council (ARC) Centre of Excellence in Quantum Biotechnology (CE230100021), ARC Future Fellowship, (FT200100401), and ARC Discovery Project (DP180101387). H.W. and S.B.S. acknowledges the support of the Ernst & Grace Matthaei Scholarship. S.B.S. acknowledges the support of the Australian Government Research Training Program Scholarship. N.P. acknowledges the support of the Melbourne Research Scholarship.

Author contributions

H.W., L.W., and A.R. conceptualized the project. H.W. performed the simulations and fabricated the device. S.B.S. built the optical setup for metasurface image processing. N.P. built the optical setup for transmission spectroscopy measurements. H.W. built the optical setup for back focal plane imaging. S.M. prepared the HeLa cell samples. W.S.L. and P.F.M.E. fabricated a preliminary version of the metasurface. H.W. and A.R. developed the initial draft of the manuscript. All authors provided their input on the manuscript. A.R., L.W., and E.H. supervised the project.

Acknowledgements

This work was performed in part at the Melbourne Centre for Nanofabrication (MCN), and the Micro Nano Research Facility at RMIT University in the Victorian Node of the Australian National Fabrication Facility (ANFF). The authors thank Neuton Li for their very helpful discussions.

# Supplementary Information: Wavelength and Polarization Multiplexed Nonlocal Metasurface for Quantitative Phase Microscopy

Haiwei Wang[1], Shikun Ma[2,3], Shaban B. Sulejman[1], Niken Priscilla[1], Wendy S.L. Lee[1], Peter Francis Matthew Elango[4,5], Lukas Wesemann[1], Elizabeth Hinde[2] and Ann Roberts[1]

[1] ARC Centre of Excellence for Transformative Meta-Optical Systems, School of Physics, The University of Melbourne, Victoria 3010, Australia

[2] School of Physics, The University of Melbourne, Victoria 3010, Australia

[3] Department of Biochemistry and Pharmacology, The University of Melbourne, Victoria 3010, Australia

[4] Functional Materials and Microsystems Research Group and the Mico Nano Research Facility, RMIT University, Melbourne, 3001, Australia

[5] ARC Centre of Excellence for Transformative Meta-Optical Systems, RMIT University, Melbourne, 3001, Australia

## Supplementary Information 1: Fourier transforms and angular spectrum

### 1.1 Two-dimensional Fourier transform

Consider a 2D position space scalar function $f(\boldsymbol{\rho})$ of the position coordinates $\boldsymbol{\rho} = (x, y)$. Its spatial frequency space representation denoted by the function $\tilde{f}(\boldsymbol{u})$ of the spatial frequency coordinates $\boldsymbol{u} = (u_x, u_y)$, is given by the Fourier transform

$$\tilde{f}(\boldsymbol{u}) = \mathcal{F}\{f(\boldsymbol{\rho})\}(\boldsymbol{u}) \equiv \int f(\boldsymbol{\rho})e^{-i2\pi \boldsymbol{u}\cdot\boldsymbol{\rho}}d^2\boldsymbol{\rho} = \int f(x,y)e^{-i2\pi(u_x x + u_y y)}dxdy, \qquad \text{(SE1)}$$

where $i$ is the imaginary unit, and $\mathcal{F}$ is the Fourier transform operator. The inverse Fourier transform recovers the original function from its Fourier transform, it is defined as

$$f(\boldsymbol{\rho}) = \mathcal{F}^{-1}\{\tilde{f}(\boldsymbol{u})\}(\boldsymbol{\rho}) \equiv \int \tilde{f}(\boldsymbol{u})e^{i2\pi \boldsymbol{u}\cdot\boldsymbol{\rho}}d^2\boldsymbol{u} = \int \tilde{f}(u_x, u_y)e^{i2\pi(u_x x + u_y y)}du_x du_y. \tag{SE2}$$

## 1.2 Angular spectrum of scalar fields

The Fourier transform formalism is used to describe the propagation of a scalar field in a homogeneous medium. A scalar field $\psi(\boldsymbol{\rho}, z)$ is written as a function of transverse coordinates $\boldsymbol{\rho} = (x, y)$, and a longitudinal coordinate $z$. We assume the scalar field is a solution of the Helmholtz equation

$$(\nabla^2 + 4\pi^2 u_0^2)\psi(\boldsymbol{\rho}, z) = 0, \tag{SE3}$$

Where $\nabla^2 = \frac{\partial^2}{\partial x^2} + \frac{\partial^2}{\partial y^2} + \frac{\partial^2}{\partial z^2}$ is the Laplace operator, and $u_0$ is the spatial frequency of the wave, given by the frequency of the oscillation divided by the wave propagation speed. Assuming that the field propagates in the $+z$ direction only, $\psi(\boldsymbol{\rho}, z)$ can be determined if $\psi(\boldsymbol{\rho}, 0)$ is known, with

$$\psi(\boldsymbol{\rho}, z) = \int \tilde{\psi}(\boldsymbol{u}, 0) e^{i2\pi \boldsymbol{u}\cdot\boldsymbol{\rho}} e^{i2\pi u_z(\boldsymbol{u}) z} d^2\boldsymbol{u}, \tag{SE4}$$

where

$$\tilde{\psi}(\boldsymbol{u}, 0) = \int \psi(\boldsymbol{\rho}, 0) e^{-i2\pi \boldsymbol{u}\cdot\boldsymbol{\rho}} d^2\boldsymbol{\rho}, \tag{SE5}$$

is the angular spectrum of the field, and

$$u_z(\boldsymbol{u}) = \sqrt{u_0^2 - |\boldsymbol{u}|^2} = \sqrt{u_0^2 - u_x^2 - u_y^2}. \tag{SE6}$$

Thus, the propagation of the field is described by a superposition of plane waves at different incident angles. $\tilde{\psi}(\boldsymbol{u}, 0)$ is called the angular spectrum since it corresponds to the amplitude of a plane wave propagating in the direction given by $\boldsymbol{s} = (u_x/u_0, u_y/u_0, u_z/u_0)$ when $|\boldsymbol{u}| \leq u_0$. For the case $|\boldsymbol{u}| > u_0$, the field no longer propagates along the $z$-direction but decays exponentially.

Diffraction of a scalar field by a thin transmission mask can be described in this framework. Given an incident field $\psi_i(\boldsymbol{\rho})$ and a thin mask $t(\boldsymbol{\rho})$, the transmitted field immediately after the mask $\psi_t(\boldsymbol{\rho})$ is given by $\psi_t(\boldsymbol{\rho}) = t(\boldsymbol{\rho})\psi_i(\boldsymbol{\rho})$. The field after the transmission mask can be determined by propagating $\psi_t(\boldsymbol{\rho})$.

## 1.3 Angular spectrum of electromagnetic fields

An electromagnetic field is specified by two vector fields: the electric field $\boldsymbol{E}(\boldsymbol{\rho}, z) = (E_x, E_y, E_z)$, and the magnetic field $\boldsymbol{H}(\boldsymbol{\rho}, z) = (H_x, H_y, H_z)$. In a homogeneous medium, each Cartesian component of the electric and magnetic field satisfies the Helmholtz equation (SE3), and each component can be propagated using SE4.

Diffraction of an electromagnetic field by a thin mask differs from that of the scalar field. This stems from the requirement that each plane wave component must be transverse to its direction of propagation. Given an incident vector field $\boldsymbol{E}_i(\boldsymbol{\rho})$, and scalar transmission mask $t(\boldsymbol{\rho})$, the transmitted field is given by $\boldsymbol{E}_t(\boldsymbol{\rho}) = t(\boldsymbol{\rho})\boldsymbol{E}_i(\boldsymbol{\rho})$. To satisfy the transverse wave condition, the angular spectrum of the transmitted field will not be the Fourier transform of each Cartesian component, it is instead[1]

$$\widetilde{\boldsymbol{E}}_t(\boldsymbol{u}) = \left(\mathbb{I} - \frac{\boldsymbol{u}\boldsymbol{u}}{u_0^2}\right)\mathcal{F}\{t(\boldsymbol{\rho})\boldsymbol{E}_i(\boldsymbol{\rho})\}(\boldsymbol{u}), \tag{SE7}$$

Where $\mathbb{I}$ is the identity tensor. In terms of Cartesian components, this is

$$\begin{pmatrix}\widetilde{E}_{t,x}(\boldsymbol{u})\\ \widetilde{E}_{t,y}(\boldsymbol{u})\\ \widetilde{E}_{t,z}(\boldsymbol{u})\end{pmatrix} = \begin{pmatrix}1-\frac{u_x^2}{u_0^2} & -\frac{u_xu_y}{u_0^2} & -\frac{u_xu_z}{u_0^2}\\ -\frac{u_xu_y}{u_0^2} & 1-\frac{u_y^2}{u_0^2} & -\frac{u_yu_z}{u_0^2}\\ -\frac{u_xu_z}{u_0^2} & -\frac{u_yu_z}{u_0^2} & 1-\frac{u_z^2}{u_0^2}\end{pmatrix}\begin{pmatrix}\mathcal{F}\{t(\boldsymbol{\rho})E_{i,x}(\boldsymbol{\rho})\}(\boldsymbol{u})\\ \mathcal{F}\{t(\boldsymbol{\rho})E_{i,y}(\boldsymbol{\rho})\}(\boldsymbol{u})\\ \mathcal{F}\{t(\boldsymbol{\rho})E_{i,z}(\boldsymbol{\rho})\}(\boldsymbol{u})\end{pmatrix}. \tag{SE8}$$

Thus, given an incident plane wave with amplitude $E_0$ and uniform polarization $\hat{\boldsymbol{e}}_{in}$ as in the main text, the diffracted angular spectrum is

$$\widetilde{\boldsymbol{E}}_t(\boldsymbol{u}) = E_0 \tilde{t}(\boldsymbol{u})\left(\mathbb{I} - \frac{\boldsymbol{u}\boldsymbol{u}}{u_0^2}\right)\hat{\boldsymbol{e}}_{in}, \tag{SE9}$$

which shows that the diffracted plane wave components will not have the same polarization as the incident polarization. However, for small diffraction angles, which is assumed in the main text, $\boldsymbol{uu}/u_0^2$ is approximately zero, then, the diffracted plane waves composing the object field will have the same polarization as the incident polarization.

# Supplementary Information 2: Derivation of the intensity transfer functions

## 2.1 Calculation of the phase transfer function

The intensity of a coherent field generated by a plane wave at normal incidence with intensity $I_0 = |E_0|^2$ illuminating an object $t(\boldsymbol{\rho})$ and after passing through an optical system described by the coherent transfer function $H(\boldsymbol{u})$ is given by

$$I(\boldsymbol{\rho}) = I_0 \left| \iint d^2\boldsymbol{u}\, H(\boldsymbol{u}) \exp(i2\pi\boldsymbol{\rho} \cdot \boldsymbol{u}) \iint d^2\boldsymbol{\rho}'\, t(\boldsymbol{\rho}') \exp(-i2\pi\boldsymbol{\rho}' \cdot \boldsymbol{u}) \right|^2. \tag{SE10}$$

Expanding the absolute value squared:

$$\begin{aligned}I(\boldsymbol{\rho}) = I_0 &\iint d^2\boldsymbol{u}\, H(\boldsymbol{u}) \exp(i2\pi\boldsymbol{\rho} \cdot \boldsymbol{u}) \iint d^2\boldsymbol{\rho}'\, t(\boldsymbol{\rho}') \exp(-i2\pi\boldsymbol{\rho}' \cdot \boldsymbol{u}) \\ &\times \iint d^2\boldsymbol{u}'\, H^*(\boldsymbol{u}') \exp(-i2\pi\boldsymbol{\rho} \cdot \boldsymbol{u}') \iint d^2\boldsymbol{\rho}''\, t^*(\boldsymbol{\rho}'') \exp(i2\pi\boldsymbol{\rho}'' \cdot \boldsymbol{u}').\end{aligned} \tag{SE11}$$

We apply the weak phase approximation, where

$$t(\boldsymbol{\rho}')t^*(\boldsymbol{\rho}'') \approx 1 + i[\phi(\boldsymbol{\rho}') - \phi(\boldsymbol{\rho}'')]. \tag{SE12}$$

We compute these three terms separately, for the first term

$$\begin{aligned}I_{\mathrm{BG}}(\boldsymbol{\rho}) = I_0 \iint \cdots \iint & d^2\boldsymbol{u}\, d^2\boldsymbol{u}'\, d^2\boldsymbol{\rho}'\, d^2\boldsymbol{\rho}''\, H(\boldsymbol{u})H^*(\boldsymbol{u}') \exp[i2\pi\boldsymbol{\rho} \cdot (\boldsymbol{u} - \boldsymbol{u}')] \\ & \times \exp(-i2\pi\boldsymbol{\rho}' \cdot \boldsymbol{u}) \exp(i2\pi\boldsymbol{\rho}'' \cdot \boldsymbol{u}'),\end{aligned} \tag{SE13}$$

$$I_{\mathrm{BG}}(\boldsymbol{\rho}) = I_0 |H(0)|^2. \tag{SE14}$$

For the second term

$$I_{\text{Ph}}(\boldsymbol{\rho}) = iI_0 \iint \cdots \iint d^2\boldsymbol{u}\, d^2\boldsymbol{u}_{\text{d}}\, d^2\boldsymbol{\rho}'\, d^2\boldsymbol{\rho}''\, H(\boldsymbol{u})H^*(\boldsymbol{u} - \boldsymbol{u}_{\text{d}})\exp(i2\pi\boldsymbol{\rho}\cdot\boldsymbol{u}_{\text{d}})$$
$$\times \exp(-i2\pi\boldsymbol{\rho}'\cdot\boldsymbol{u})\exp[i2\pi\boldsymbol{\rho}''\cdot(\boldsymbol{u}-\boldsymbol{u}_{\text{d}})][\phi(\boldsymbol{\rho}') - \phi(\boldsymbol{\rho}'')],$$
(SE15)

where we transformed the integration variable $\boldsymbol{u}'$ to $\boldsymbol{u}_{\text{d}} = \boldsymbol{u} - \boldsymbol{u}'$. Integrating over $\boldsymbol{u}$ and using the fact that $\phi(\boldsymbol{\rho})$ is a real function so $\tilde{\phi}^*(\boldsymbol{u}) = \tilde{\phi}(-\boldsymbol{u})$

$$I_{\text{Ph}}(\boldsymbol{\rho}) = iI_0 \iint d^2\boldsymbol{u}_{\text{d}}\, \exp(i2\pi\boldsymbol{\rho}\cdot\boldsymbol{u}_{\text{d}})[H(\boldsymbol{u}_{\text{d}})H^*(0) - H(0)H^*(-\boldsymbol{u}_{\text{d}})]\tilde{\phi}(\boldsymbol{u}_{\text{d}}).$$
(SE16)

Therefore, we see that the Fourier spectrum of the intensity can be written as

$$\tilde{I}(\boldsymbol{u}) = I_0|H(0)|^2\delta^2(\boldsymbol{u}) + H_\phi(\boldsymbol{u})\tilde{\phi}(\boldsymbol{u}),$$
(SE17)

where

$$H_\phi(\boldsymbol{u}) = iI_0[H^*(0)H(\boldsymbol{u}_{\text{d}}) - H(0)H^*(-\boldsymbol{u}_{\text{d}})]$$
(SE18)

is the phase transfer function.

## 2.2 Calculation of the DPC transfer function

The DPC image is calculated using two images $I_+(\boldsymbol{\rho})$ and $I_-(\boldsymbol{\rho})$ filtered with CTFs $H_+(\boldsymbol{u})$ and $H_-(\boldsymbol{u})$, respectively. The sum of the Fourier spectra of the intensities is

$$\tilde{I}_+(\boldsymbol{u}) + \tilde{I}_-(\boldsymbol{u}) = I_0(|H_+(0)|^2 + |H_-(0)|^2)\delta^2(\boldsymbol{u}) + [H_{\phi,+}(\boldsymbol{u}) + H_{\phi,-}(\boldsymbol{u})]\tilde{\phi}(\boldsymbol{u}).$$
(SE19)

When $H_{\phi,+}(\boldsymbol{u}) = -H_{\phi,-}(\boldsymbol{u})$ is satisfied, the sum of the images will not contain any phase contrast information. Note that this condition does not require the CTFs $H_+(\boldsymbol{u})$ and $H_-(\boldsymbol{u})$ to be perfect mirror images of each other. The difference of the Fourier spectra of the intensities is

$$\tilde{I}_+(\boldsymbol{u}) - \tilde{I}_-(\boldsymbol{u}) = I_0(|H_+(0)|^2 - |H_-(0)|^2)\delta^2(\boldsymbol{u}) + [H_{\phi,+}(\boldsymbol{u}) - H_{\phi,-}(\boldsymbol{u})]\tilde{\phi}(\boldsymbol{u}).$$
(SE20)

The Fourier spectrum of the DPC intensity calculated using Eq. 4 in the main text is

$$\tilde{I}_{\text{DPC}}(\boldsymbol{u}) = \frac{|H_+(0)|^2 - |H_-(0)|^2}{|H_+(0)|^2 + |H_-(0)|^2}\delta^2(\boldsymbol{u}) + \frac{H_{\phi,+}(\boldsymbol{u}) - H_{\phi,-}(\boldsymbol{u})}{|H_+(0)|^2 + |H_-(0)|^2}\tilde{\phi}(\boldsymbol{u}).$$
(SE21)

The first term is a DC offset, which is physically irrelevant in phase retrieval and can be neglected, while the second term contains the phase information. The DPC transfer function in Eq. 5 of the main text is hence given by

$$H_{\mathrm{DPC}}(\boldsymbol{u}) = \frac{H_{\phi,+}(\boldsymbol{u}) - H_{\phi,-}(\boldsymbol{u})}{|H_+(0)|^2 + |H_-(0)|^2}. \tag{SE22}$$

# Supplementary Information 3: Additional metasurface characterization data

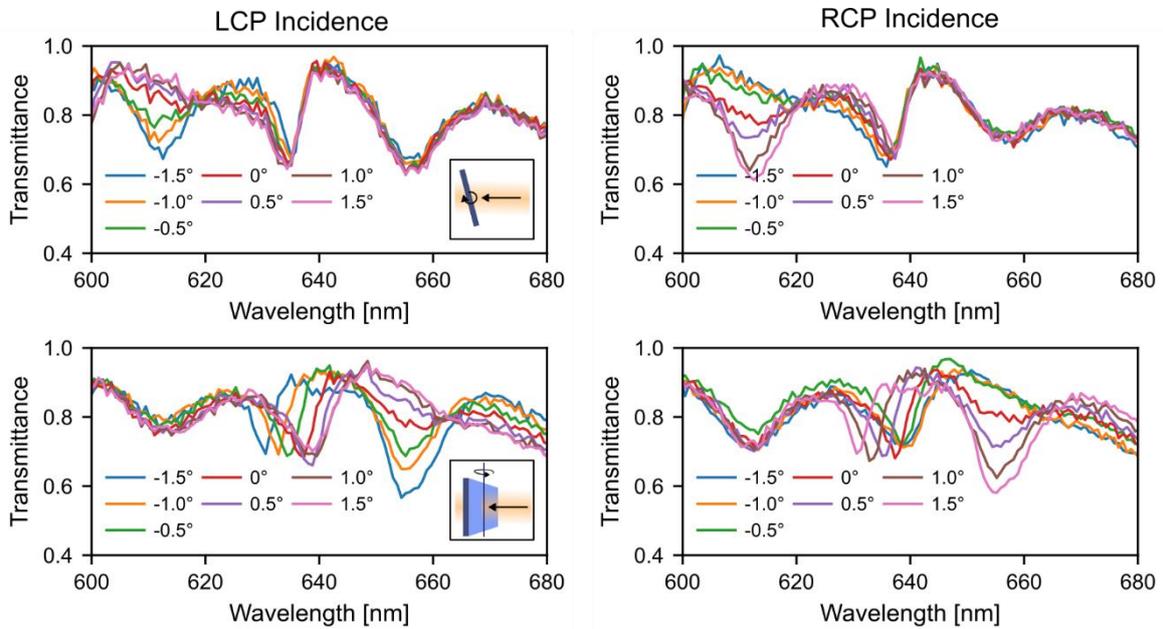

**Supplementary Figure 1: Transmission spectra measurements at varying incident angles for LCP vs. RCP incident light.**

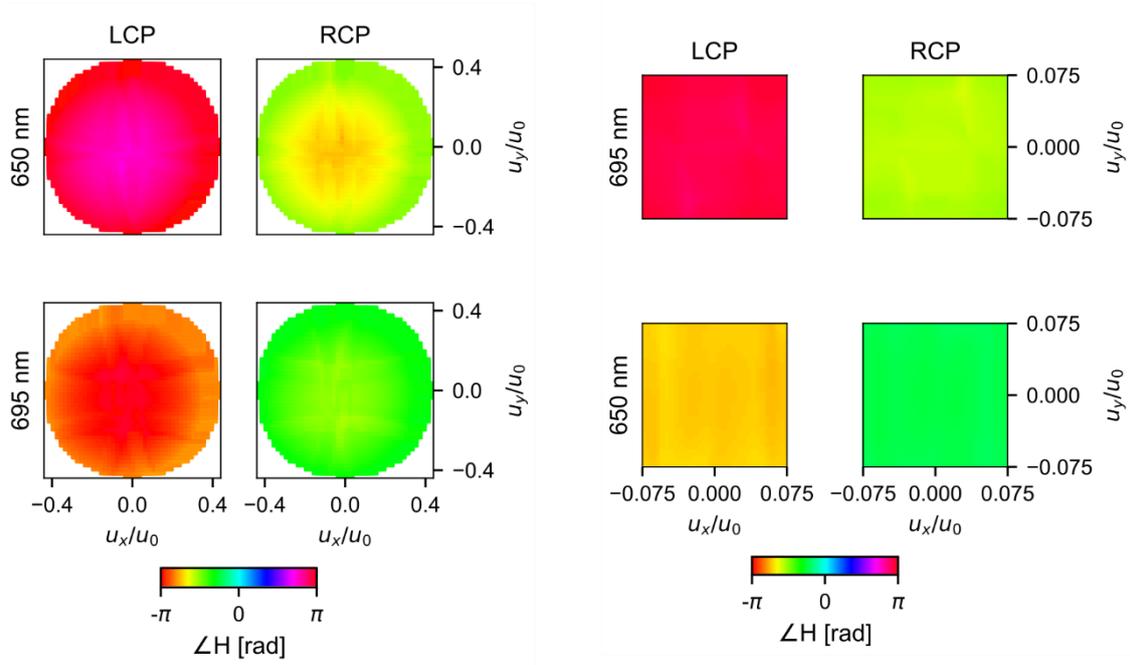

**Supplementary Figure 2: Numerically calculated phase of the coherent transfer function.**

# Supplementary Information 4: Calibrating the phase response of the SLM

The Jones matrix of SLM is expressed as

$$M(\phi) = \begin{pmatrix} e^{i\phi} & 0 \\ 0 & 1 \end{pmatrix}, \tag{SE23}$$

where the first component is horizontal polarization, the horizontal plane corresponds to the plane of rotation of the liquid crystal molecules, hence a variable, voltage-dependent phase shift is applied to this component

$$\phi(V) = \frac{2\pi[n_e(V) - n_o](2d)}{\lambda_0}, \tag{SE24}$$

where $n_e(V)$ is the variable extraordinary refractive index, $n_o$ is the ordinary refractive index, $d$ is the thickness of the liquid crystal layer, and $\lambda_0$ is the vacuum wavelength. A factor of two multiplies the thickness because light must make a round trip through the liquid crystal layer to be reflected in a reflection-mode SLM.

With input diagonal polarization given by the Jones vector

$$\frac{1}{\sqrt{2}}\begin{pmatrix}1\\1\end{pmatrix},$$

The output polarization is

$$\frac{1}{\sqrt{2}}\begin{pmatrix}e^{i\phi}\\1\end{pmatrix}.$$

The dot product with the perpendicular diagonal polarization is

$$\frac{1}{2}(e^{i\phi}-1).$$

The resulting intensity is the modulus-squared of this

$$\frac{I(\phi)}{I_0}=\frac{1}{4}(e^{i\phi}-1)(e^{-i\phi}-1)=\frac{1}{2}(1-\cos\phi)=\sin^2\frac{\phi}{2}. \tag{SE25}$$

$I_0$ is measured as the maximum intensity value for all pixel values displayed. Using the intensity modulation measurements, we calibrate the phase response. Phase unwrapping was applied to obtain phase values above $\pi$.

Supplementary Figure 3 shows the measured phase for different pixel values displayed on the SLM obtained with two different wavelengths. The phase response is most linear at the middle of the range. We see that the slope of the curve for 610 nm is slightly steeper than 656 nm, indicating that light at 610 nm experience stronger phase shifts. By applying the wavelength correction factor according to Eq. 2 of the main text and adding a constant offset, the corrected phase values at 610 nm match well with that at 656 nm.

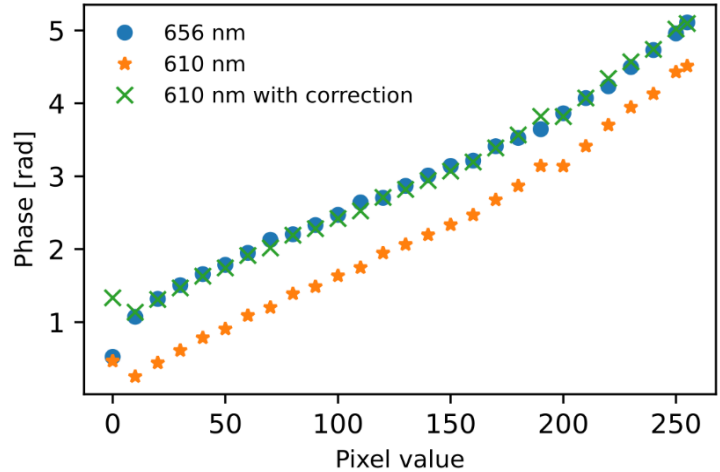

**Supplementary Figure 3: SLM phase calibration curve.**

## Supplementary Information 5: Description of optical setups

The spectroscopy setup is depicted in Supplementary Figure 4. A fiber-coupled output from an OceanOptics HL-2000-HP halogen lamp was collimated using an Olympus UPlanF 4x 0.13NA objective (OL1), producing a beam with a diameter around 1cm. Before illuminating the metasurface, a Thorlabs LPVISE100-A linear polarizer (LP) and a Thorlabs AQWP10M-580 achromatic quarter waveplate (QWP) were placed with the fast axis of the QWP oriented at $45º$ to the polarization axis of the LP to create circular polarization. The circularly polarized light then illuminated the sample which was mounted on a rotation stage to measure the spectra at varying incident angles. After the sample, a 50x Nikon CFI60 TU Plan 0.55NA objective (OL2) with a long working distance of 11 mm was used to magnify the metasurface area - the relatively long working distance was required to permit rotation of the sample. A Thorlabs LA1131-A lens (L1) with a focal length of 50 mm was placed to collimate the light collected by the objective. The image of the sample was conjugated onto an aperture to limit the subsequently collected light to only that transmitted through the metasurface. The light was then split equally into two paths by a Thorlabs CM1BS013 non-polarizing beamsplitter cube (BS), light from one path was imaged onto a Thorlabs DCC1545M CMOS camera using (L3) to allow for sample alignment, and the light in the other path was coupled into a multimode fiber (ThorLabs M14L10) to an OceanOptics HR2000+ES visible light spectrometer using L1.

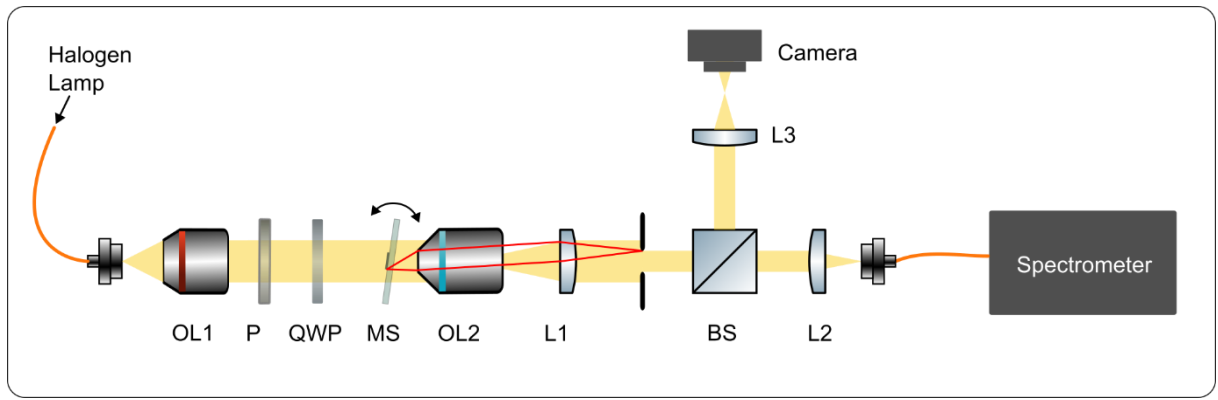

**Supplementary Figure 4: Diagram of the spectroscopy setup.**

Supplementary Figure 5 shows the diagram of the back focal plane imaging setup. The light was coupled through a single-mode fiber and collimated using an Olympus 4x 0.1NA objective (OL1). A polarizer (P, Thorlabs LPVIS050-MP2) was placed to set the incident polarization to be linearly polarized. The collimated beam illuminates the back aperture of an objective (OL2, Olympus Plan 20x 0.4NA) to create a focal spot on the metasurface. After passing through the metasurface, the light is then collected by an imaging objective lens (OL3, Olympus Plan 20x 0.4NA). The back focal plane of the imaging objective was then relayed to a polarization-sensitive camera (Thorlabs CS505MUP1) using a series of lenses (L2, L3 and L4). L2 is a biconvex 125 mm lens (Thorlabs LB1106-B), L3 is a 100 mm plano-convex lens (Thorlabs LA1509-A), and L4 is a 75 mm plano-convex lens (Thorlabs LA1608-A). A quarter waveplate (QWP, Thorlabs AHWP10M-580) is used such that circular polarizations can be distinguished by the polarization-sensitive camera instead of linear polarizations. The aperture seen in the camera corresponds to the numerical aperture of the imaging objective, given that the cone of light from the condenser has a numerical aperture greater than or equal to the imaging objective numerical aperture. This was used to correlate the pixel positions of captured images with incident directions, with normal incidence being at the center of the aperture. By sweeping through the illumination wavelength in small steps, we observed the evolution of the angular response to ascertain the wavelengths where the desired transfer functions occur.

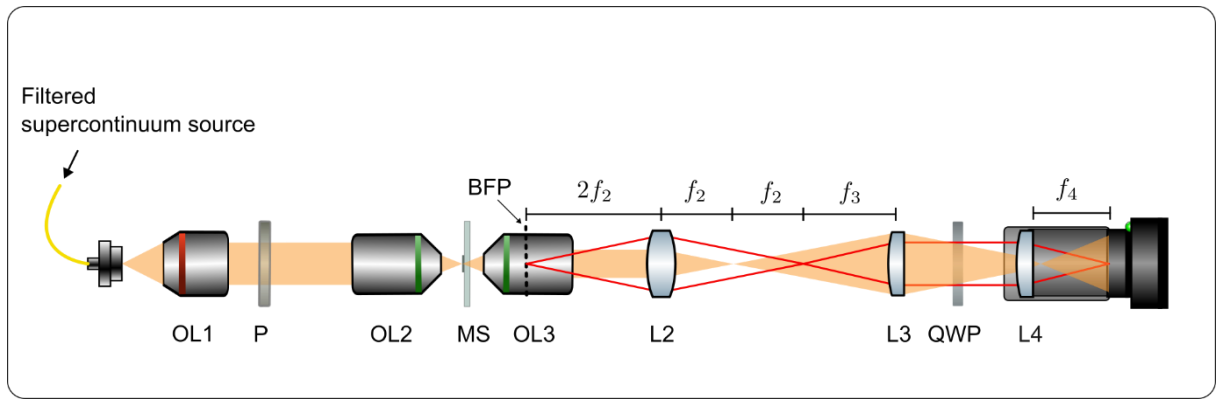

**Supplementary Figure 5: Diagram of the back focal plane imaging setup.**

Supplementary Figure 6 shows the diagram of the setup used to demonstrate phase imaging. The beam from the output collimator of the fiber-coupled filtered supercontinuum laser (NKT Photonics SuperK Compact with a SuperK Select acousto-optic tunable filter) was expanded using OL1 (Olympus Plan 20x 0.4NA objective) and L1 (Thorlabs LA1986-B) to illuminate the active area of the SLM (Holoeye Pluto 2 phase-only spatial light modulator). P is a linear polarizer (Thorlabs LPVIS050-MP2) used to set the polarization to horizontal polarization. The SLM was illuminated with horizontally polarized light, such that pure phase modulations are achieved without altering the output polarization state. The SLM was mounted on a two-axis rotation stage to control the tip and tilt angle for adjustment of the incident angle of the beam onto the metasurface downstream of the setup. The second lens system consisting of L2 (Thorlabs LA1433-B) and OL2 (Olympus Plan 20x 0.4NA objective) then demagnifies the field reflected from the SLM onto the metasurface. This image then acts as the object. After passing through the metasurface, OL3 (Nikon TU Plan ELWD 20x 0.4NA objective) collects the processed image and L3 (Thorlabs LA1433-A) images. A quarter waveplate (QWP) (Thorlabs AHWP10M-580) is placed before L3 with its fast axis oriented at $45°$ with respect to the $x$-axis such that the circular polarization components of the light are converted into horizontal and linear polarizations to be analyzed by a polarization-sensitive camera.

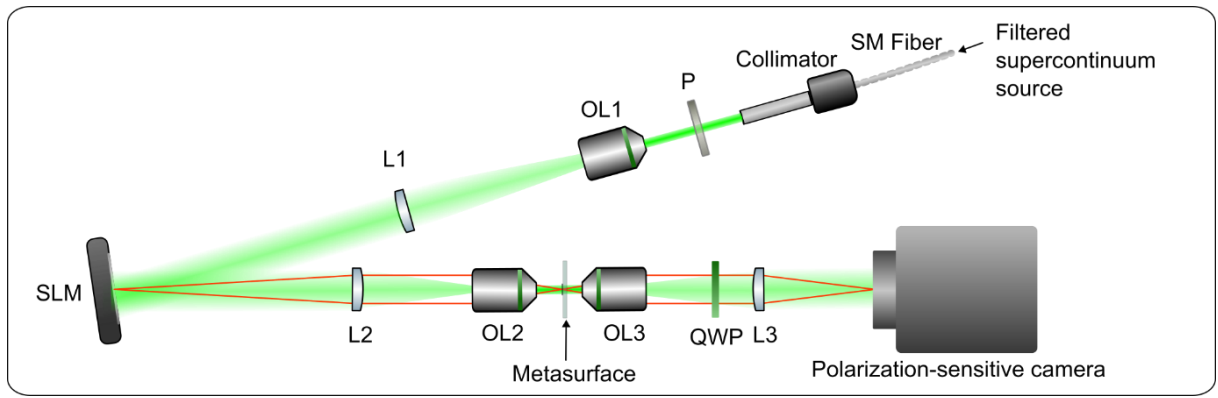

**Supplementary Figure 6: Diagram of the phase imaging testbed.**

References

1. Song, C., He, J. & Yuan, G. Generic full-vector angular spectrum method for calculating diffraction of arbitrary electromagnetic fields. *J. Phys. Photonics* **7**, 045021 (2025).